\documentclass[pdftex,11pt]{article}
\usepackage[utf8]{inputenc}

\usepackage[dvips]{graphicx}
\usepackage{pslatex}

\usepackage{amssymb}
\usepackage{gensymb}
\usepackage{amsmath}
\usepackage{enumitem}

\usepackage{sectsty}
\sectionfont{\fontsize{14}{12}\selectfont}
\subsectionfont{\fontsize{12}{12}\selectfont}

\usepackage[usenames,dvipsnames]{xcolor}
\usepackage{natbib}
\usepackage{hyperref}
\hypersetup{colorlinks,citecolor=MidnightBlue,linkcolor=red,urlcolor=black}

\usepackage[left=1in,right=1in,top=0.9in,bottom=1.1in,footskip=0.5in]{geometry}
\usepackage[onehalfspacing]{setspace}

\usepackage{xurl}

\begin{document}

\onehalfspacing

\centerline{\textbf{\Large{Observations of ubiquitous nighttime temperature inversions}}}
\centerline{\textbf{\Large{in Mars' tropics after large-scale dust storms}}}
\vspace{2.0em}
\centerline{L. J. Steele, A. Kleinb\"{o}hl and D. M. Kass}
\vspace{2.0em}
\centerline{Jet Propulsion Laboratory, California Institute of Technology, Pasadena, California, USA.} 
\centerline{Corresponding author: Liam Steele (liam.j.steele@jpl.nasa.gov)}
\vspace{2.0em}
\hrulefill


\vspace{2.0em}
\section*{Key points}


\begin{itemize}[label=\raisebox{0.35ex}{\scriptsize$\bullet$}]
  \item The ubiquitous occurrence of nighttime temperature inversions in the tropical martian atmosphere during the dusty season is reported.
  \item Temperature inversions form in the early hours of the morning during the decay phase of large-scale regional dust storms.
  \item Inversions form due to a combination of cloud radiative cooling and tides excited by topography and dust abundance.
\end{itemize}


\newpage
\section*{Abstract}

We report the ubiquitous occurrence of nighttime temperature inversions in the tropical martian atmosphere during the dusty season, as observed by the Mars Climate Sounder. The inversions are linked to the occurrence of large-scale regional dust storms, with their strengths largely correlated to the strengths of the dust storms. Inversions strengthen between 2 am and 4 am, with the bases of the inversions getting cooler, and the tops of the inversions getting warmer. The inversions are strongest around Tharsis and Terra Sabaea, which are higher-elevation regions, suggesting they are forming due to a combination of topographically-excited tides and cloud radiative cooling. However, inversions are also observed over the flat plains, and are likely associated with stronger tides resulting from the increased dust abundance. These results highlight an important interplay between the dust distribution, water ice clouds and thermal tides.


\section*{Plain Language Summary}

The Mars Climate Sounder is an instrument orbiting Mars that measures vertical profiles of temperature, dust and water ice. Here we use data covering six different Mars years (obtained between December 2007 and March 2019) to investigate the nighttime atmosphere at times of year when large dust storms occur. As the dust storms begin to decay, we find that strong temperature inversions (where the temperature increases with height, which is the reverse of the typical behavior) form in the early hours of the morning. These inversions typically occur in two mountainous regions of the planet, but we also observe them commonly over the flat plains, which is unexpected. We suggest that the inversions are forming due to a combination of atmospheric temperature variations, caused by the diurnal variation of incoming sunlight over the course of a martian day, and the radiative effects of dust and water ice clouds. Our results show how varied the nighttime atmosphere of Mars in the dusty season can be, and how there is an important interplay between the dust distribution, water ice clouds and thermal tides which requires further investigation.


\newpage

\section{Introduction}

Temperature inversions are regions of the atmosphere where the temperature increases with altitude, which is the reverse of the typical behavior in the lower and middle atmosphere of Mars, away from the polar vortices \citep{McCleese2008, McCleese2010}. Temperature inversions can reveal important details about the local and large-scale dynamics of the atmosphere, microphysical process, and the radiative impact of aerosols.

Temperature inversions in the middle atmosphere of Mars (at altitudes above $\sim$50 km) were first revealed in detail by the entry profiles of the Viking landers, and linked to heating associated with the diurnal tide \citep{Seiff1977}. Just over 21 years later, the entry profile from Mars Pathfinder revealed similar wave features, but also a strong temperature inversion in the lower atmosphere, between 8--15 km \citep{Magalhaes1999}. These entry profiles span the range \textit{L}\textsubscript{\scriptsize{S}} = 96--142\degree{}, which is during Mars' aphelion season, when dust opacity is low, water ice opacity in the tropics is high, and there is little interannual variability in the atmosphere \citep{Smith2004, McCleese2010}. One-dimensional microphysics models and global climate models revealed that the temperature inversion was likely the result of nighttime cooling by water ice clouds \citep{Colaprete1999, Haberle1999, Colaprete2000}.

More extensive (though still spatially and seasonally limited) sets of profiles displaying temperature inversions were obtained via radio occultation (RO) measurements. Again, these data covered the aphelion season, with the Mars Global Surveyor RO profiles spanning \textit{L}\textsubscript{\scriptsize{S}} = 134--162\degree{} \citep{Hinson2004} and the Mars Reconnaissance Orbiter RO profiles covering two periods around \textit{L}\textsubscript{\scriptsize{S}} = 35\degree{} and \textit{L}\textsubscript{\scriptsize{S}} = 120\degree{} \citep{Hinson2014}. The strongest and most frequent inversions were found to correspond to regions of elevated terrain, particularly Tharsis, with only a few inversions over the plains. Numerical simulations suggest that the temperature inversions are the result of thermal tides, amplified by the radiative cooling of water ice clouds, with convective instability induced below the clouds \citep{Hinson2004, Hinson2014, Wilson2014, Spiga2017}. These temperature inversions were found to be a common feature of the aphelion season in analysis of multiple years of Mars Climate Sounder (MCS) data \citep{Wilson2014}.

Analyses of temperature profiles by the Mars Climate Sounder (MCS) suggest that temperature inversions can also occur in the tropics during southern hemisphere spring equinox, where they are found at altitudes $\sim$15--25 km higher than those seen in the aphelion season RO observations \citep{Steele2021}. Here, we analyze six Mars years of MCS data, and make use of the ability of the MCS instrument to observe the atmosphere at different local times. This allows us to study the evolution of temperature inversions, and their relationship with dust and water ice aerosols, with local time. We show that these inversions are prominent recurring features in the martian atmosphere that are observed in every Mars year studied, and are linked to the occurrence of large-scale dust storms.

\section{Mars Climate Sounder Data}

\begin{figure}[t]
  \includegraphics[width=1.0\linewidth]{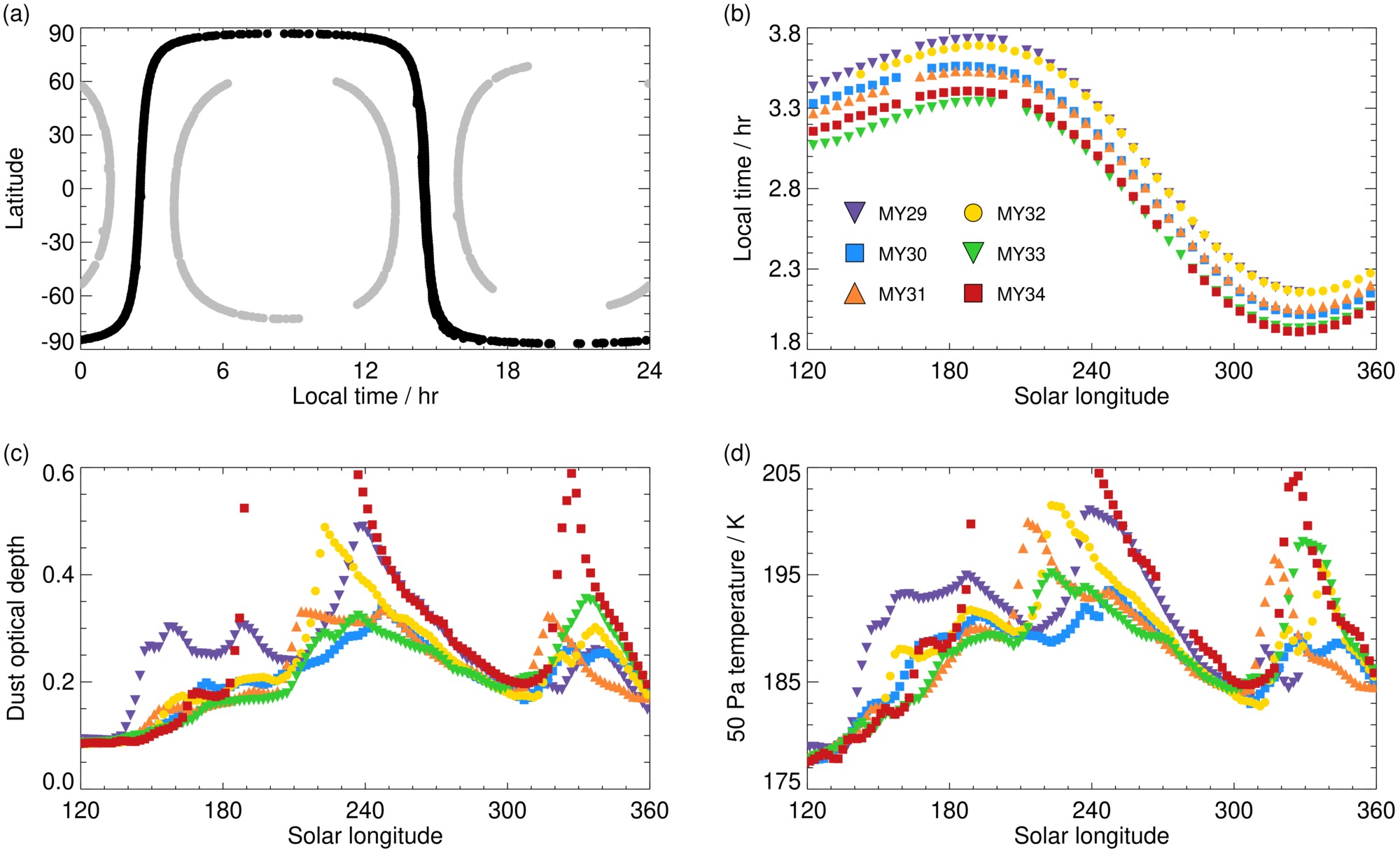}
  \caption{(a) Local time coverage of in-track (black) and cross-track (grey) measurements. Data are from \textit{L}\textsubscript{\scriptsize{S}} = 270--275\degree{}, MY33. (b) Seasonal variation of the local equator-crossing times from MY29--34 nighttime in-track measurements. (c) Seasonal variation of the 9.3 $\mu$m absorption dust optical depth, normalized to 610 Pa \citep{Montabone2015, Montabone2020}. (d) Seasonal variation of the MCS 50 Pa nighttime temperature. Data in (c) and (d) are zonally averaged between $\pm$15\degree{} latitude.}
  \label{fig:fig_loct}
\end{figure}

The MCS instrument is a nine-channel limb radiometer on board the Mars Reconnaissance Orbiter, with 8 mid- and far-infrared (IR) channels and one visible/near-IR channel. Each channel consists of 21 detectors, providing a radiance profile from the surface to $\sim$90 km, with $\sim$5 km (half a scale height) resolution \citep{McCleese2007}. The retrieval algorithm inverts the radiance profiles to produce geophysical quantities, including vertical profiles of temperature, dust and water ice extinction \citep{Kleinbohl2009, Kleinbohl2011, Kleinbohl2017}. MCS typically observes the limb in the direction of MRO's orbital movement, producing in-track measurements. However, beginning in Mars year (MY) 30, and continuing through MY35, the observation strategy was modified to also view the limb at 90\degree{} to the orbit trajectory, producing cross-track measurements at additional local times \citep{Kleinbohl2013}. Figure \ref{fig:fig_loct}a shows how the local time of the measurements varies with latitude. As the orbit moves from the equator to $\pm$60\degree{} latitude, the time difference between the in-track and cross-track measurements increases from $\sim$1.3 hours to $\sim$3 hours. Cross-track measurements were taken almost continually between \textit{L}\textsubscript{\scriptsize{S}} = 225\degree{}, MY32 and \textit{L}\textsubscript{\scriptsize{S}} = 360\degree{}, MY33. Outside of this period, the cross-track observations occurred in campaigns of around 4 weeks duration \citep{Kleinbohl2013}.

The local times of the measurements also vary with season and Mars year. Figure \ref{fig:fig_loct}b shows how the local time of the nighttime in-track measurements varies between \textit{L}\textsubscript{\scriptsize{S}} = 120--360\degree{} in MY29--34. Local times are at their latest value for the year around \textit{L}\textsubscript{\scriptsize{S}} = 190\degree{}, and this is also when the largest time difference between years, of $\sim$25 minutes, occurs. Local times are at their earliest value for the year around \textit{L}\textsubscript{\scriptsize{S}} = 330\degree{}. For simplicity, and as we are focusing on the tropics, when referring to local times we will reference the equator-crossing local time. For the analysis here we binned the data by 10\degree{} in longitude, 5\degree{} in latitude and 5\degree{} in \textit{L}\textsubscript{\scriptsize{S}}. We only show nighttime data, as this is when the temperature inversions are seen.

\section{Overview of Dust Storms}

The largest interannual variation in the martian climate is due to variations in the dustiness of the atmosphere, particularly during the dust storm season of southern hemisphere spring and summer (\textit{L}\textsubscript{\scriptsize{S}} = 180--360\degree). All Mars years display a mixture of local and regional dust storms \citep{Montabone2015, Wang2015, Kass2016}, with large-scale regional storms labeled A, B or C, depending on when they form \citep{Kass2016}. The A and C storms typically originate in the northern hemisphere, and travel across the equator through the Acidalia and Utopia storm tracks \citep{Hollingsworth1996, Wang2005, Wang2015}. B storms are cap-edge events, confined to southern hemisphere high latitudes. Some years, most recently MY25, MY28 and MY34, also experience Global Dust Events (GDEs) where a thick dust haze covers a large fraction of the planet, and the whole atmosphere is affected by the dust \citep{Kass2019, Kleinbohl2020, Shirley2020}. MY29 experienced early dust storm activity, with one storm originating in the northern hemisphere at \textit{L}\textsubscript{\scriptsize{S}} $\approx$ 143\degree{}, and the other originating in the southern hemisphere at \textit{L}\textsubscript{\scriptsize{S}} $\approx$ 152\degree{} \citep{Wang2015}. 

Figure \ref{fig:fig_loct}c shows the seasonal variation of the dust optical depth in the tropics for MY29--34, from the data of \cite{Montabone2015} and \cite{Montabone2020}, while Figure \ref{fig:fig_loct}d shows the corresponding 50 Pa nighttime temperatures from MCS. It is clear that the interannual temperature variations are related to differences in dust storm behavior. The early dust storm activity in MY29 resulted in warmer and dustier tropics between \textit{L}\textsubscript{\scriptsize{S}} $\approx$ 140--200\degree{}. The interannual differences seen between \textit{L}\textsubscript{\scriptsize{S}} $\approx$ 210--240\degree{} and \textit{L}\textsubscript{\scriptsize{S}} $\approx$ 300--340\degree{} are due to differences between the A and C storms respectively. As well as directly warming the tropics, the A and C storms also produce high-latitude warming in the northern hemisphere due to a strengthened overturning circulation. B storms, which start at \textit{L}\textsubscript{\scriptsize{S}} $\approx$ 245--260\degree{}, have a smaller influence on the tropical atmosphere and lack a dynamical response in the northern hemisphere \citep{Kass2016} and hence are not considered in this study.

\section{Results}

\begin{figure}[t]
  \includegraphics[width=1.0\linewidth]{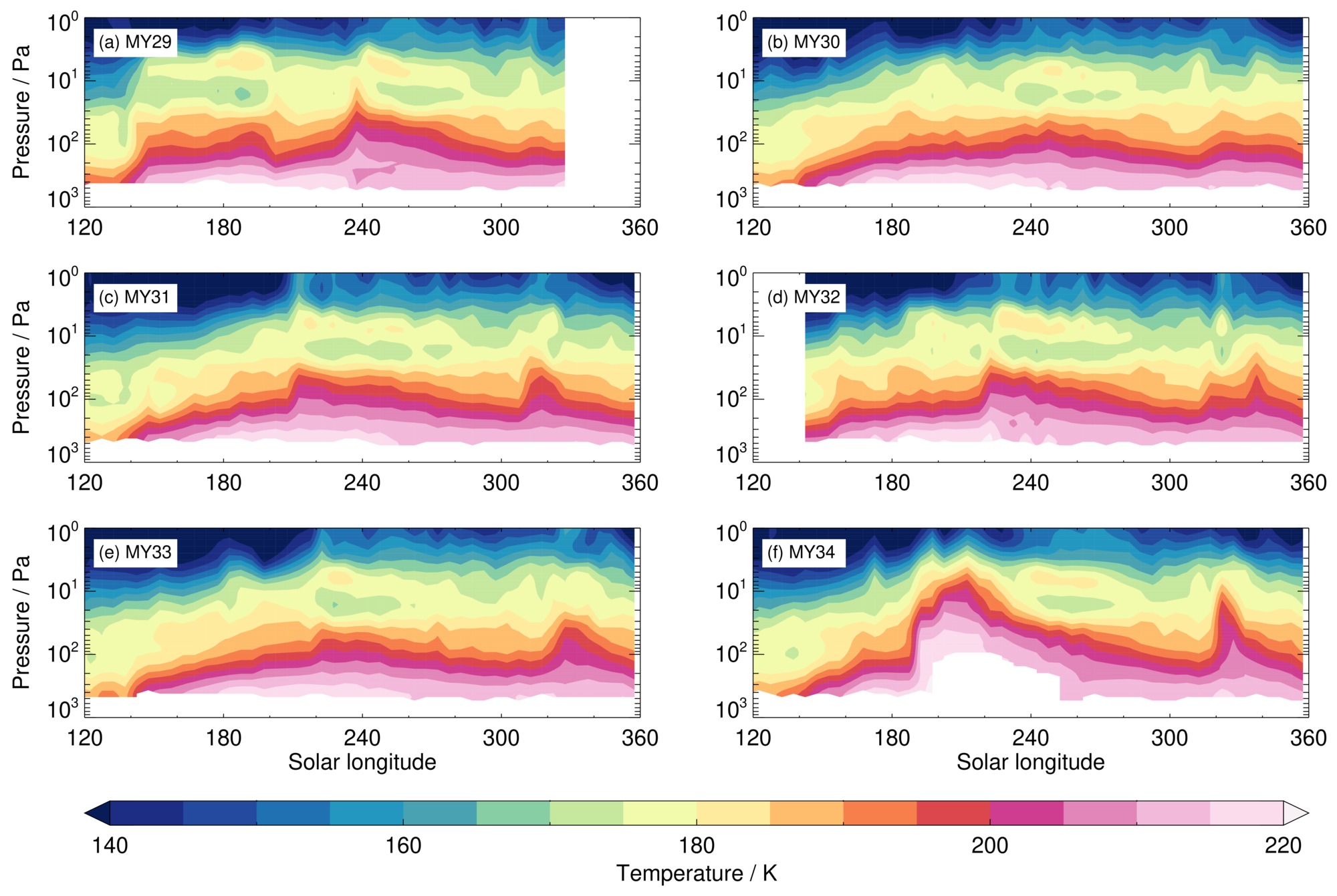}
  \caption{Nighttime temperatures, as a function of \textit{L}\textsubscript{\scriptsize{S}} and pressure, from MY29--34 in-track measurements. Data are averaged between 0--10\degree{}N and 60--90\degree{}W.}
  \label{fig:temp_6my}
\end{figure}

The seasonal evolution of nighttime temperatures in the tropics during the dusty season is shown in Figure \ref{fig:temp_6my} for MY29--34. Here we show the temperatures averaged between 0--10\degree{}N and 60--90\degree{}W, as this is the region where temperature inversions are typically strongest and longest lasting \citep{Steele2021}. As can be seen by comparing Figure \ref{fig:temp_6my} with the tropical dust column opacity in Figure \ref{fig:fig_loct}c, temperature inversions are linked to the occurrence of dust storms. In all years, inversions typically occur $\sim$5--10\degree{} of \textit{L}\textsubscript{\scriptsize{S}} after the peak of the A storms, where here we define the dust storm peak as the period with the warmest 50 Pa temperatures \citep{Kass2016}. The strengths of the inversions appear to be related to the strengths of the A storms. For example, MY30 had a very weak A storm \citep{Montabone2015, Wang2015}, and shows relatively weak inversions (Figure \ref{fig:temp_6my}b), while MY29 and MY32 had some of the strongest A storms, and show the strongest inversions at this time (Figure \ref{fig:temp_6my}a,d). Inversions also appear after the MY31 and MY32 C storms (Figure \ref{fig:temp_6my}c,d). They do not appear in MY30, which had a weak C storm, nor in MY34, which had a strong C storm (Figure \ref{fig:fig_loct}c).

Interestingly, while MY34 experienced a GDE (Figure \ref{fig:temp_6my}f), resulting in 50 Pa temperatures over 20 K warmer than at the peak of the A storms, the inversions which formed afterwards are no stronger than those that form after the A storms. They were slower in appearing though, not forming until $\sim$30\degree{} of \textit{L}\textsubscript{\scriptsize{S}} after the peak of the GDE. In fact, the strongest temperature inversions observed in the tropics by MCS occurred after the early Z storm in MY29. As seen in Figure \ref{fig:temp_6my}a, the inversions began forming at \textit{L}\textsubscript{\scriptsize{S}} $\approx$ 150\degree{}, and continued to strengthen until \textit{L}\textsubscript{\scriptsize{S}} $\approx$ 200\degree{}, at which point they rapidly disappeared. This strengthening occurred despite the active dust lifting ending by \textit{L}\textsubscript{\scriptsize{S}} $\approx$ 157\degree{} \citep{Wang2015}. This behavior is different to that after the A storms, C storms and GDE, where inversions are initially strong and then slowly weaken. This suggests that there are other factors involved in determining the strength and behavior of the temperature inversions, as opposed to just dust storm strength.

One issue when analyzing the seasonal evolution of temperature inversions is the local time of the MCS observations. As shown in Figure \ref{fig:fig_loct}b, the equator-crossing local time rapidly moves to earlier values between \textit{L}\textsubscript{\scriptsize{S}} = 210--330\degree{}. Thus, data towards the end of the year correspond to local times of $\sim$2:00 am/pm, compared to 3:30 am/pm at \textit{L}\textsubscript{\scriptsize{S}} = 180\degree{}. As has been shown in modeling studies \citep{Colaprete2000, Wilson2014}, the radiative cooling from clouds at night can have notable effects on the temperature structure over short timescales. As such, some of the weakening of the inversions seen in Figure \ref{fig:temp_6my} as \textit{L}\textsubscript{\scriptsize{S}} processes may be related to changes in the observation time, rather than real changes in the atmosphere. 

\begin{figure}[t]
  \includegraphics[width=1.0\linewidth]{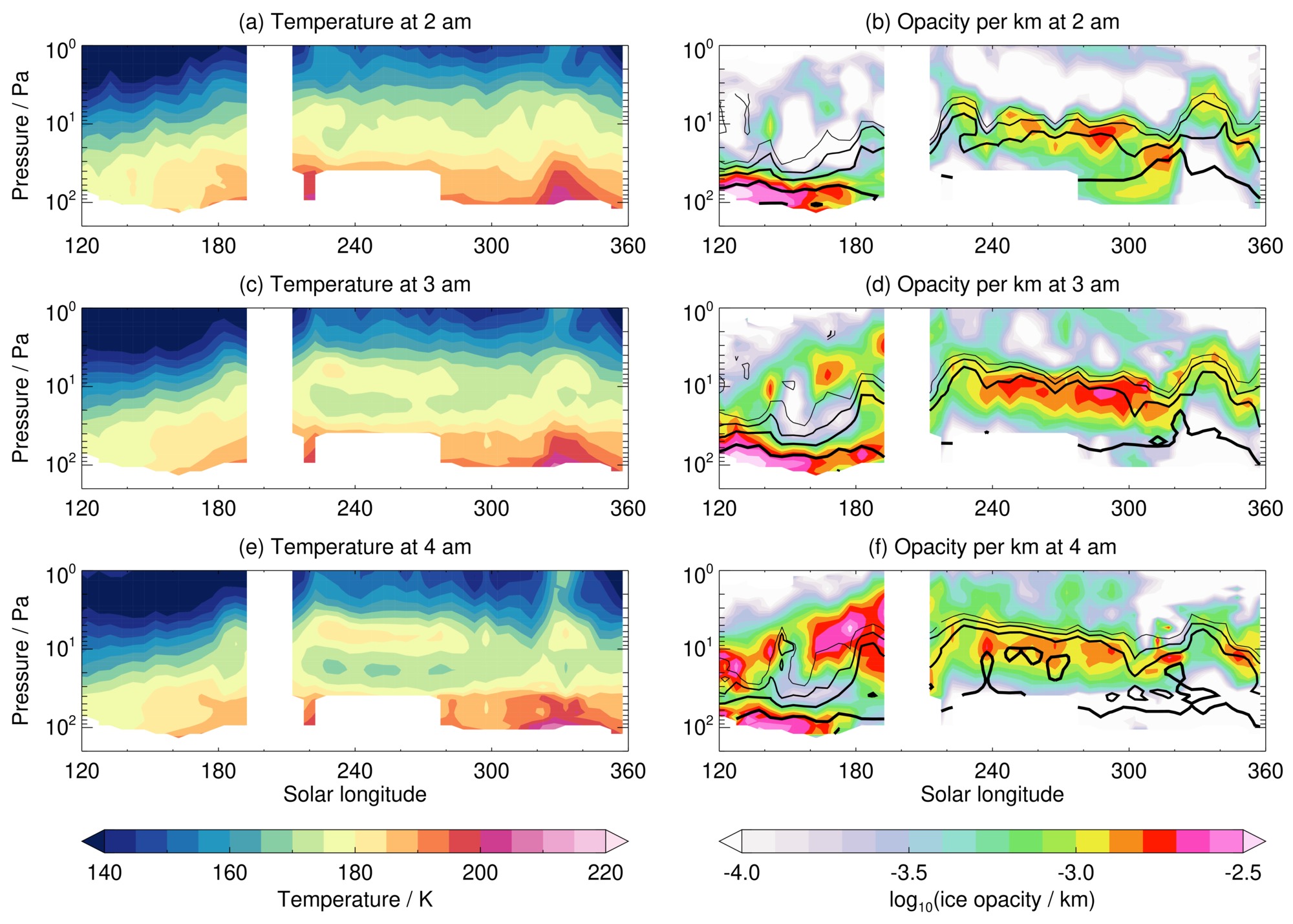}
  \caption{Nighttime temperatures (left column) and aerosol opacities (right column), as a function of \textit{L}\textsubscript{\scriptsize{S}} and pressure, at three different local times in MY33. Data are averaged between 0--10\degree{}N and 60--90\degree{}W, with white strips showing regions where no cross-track data are available. Water ice opacities are shaded, and dust opacities are represented by black contours ranging from $-$4.5 (thin black line) to $-$3 (thick black line), in steps of 0.5. Thus, the dust opacity triples with each successive contour.}
  \label{fig:tdi_ls_alt}
\end{figure}

To remove the local time effect, we use the in-track and cross-track observations from MCS, which span a period of $\sim$3 hours at the equator (see Figure \ref{fig:fig_loct}a). We use quadratic interpolation between local times to convert the data to values at constant times of 2 am, 3 am and 4 am. This interpolation is only applied if all three data points are available (one from the in-track data, and two from the cross-track data on either side). As cross-track measurements were not made continuously, and not at all in MY29, the coverage is not as complete as the in-track data used in Figure \ref{fig:temp_6my}. MY33 had the best cross-track coverage, and thus we show the results from this year in Figure \ref{fig:tdi_ls_alt}. (Figures S1--S3 in the supporting information show similar plots for MY30--32.)

Comparing the temperature structure between the in-track data (Figure \ref{fig:temp_6my}e) and the constant local time data (Figure \ref{fig:tdi_ls_alt}a,c,e), it is clear that the weakening and eventual disappearance of the inversions after the A storm in Figure \ref{fig:temp_6my}e (between \textit{L}\textsubscript{\scriptsize{S}} = 220--260\degree{}), and their lowering in the atmosphere, is an artifact caused by the observations regressing to an earlier local time. Instead, if the local time is constant, the inversions can be seen to last much longer, until \textit{L}\textsubscript{\scriptsize{S}} $\approx$ 300\degree{}. The inversions also increase in strength between 2 am and 4 am, with the bases of the inversions getting cooler, and the tops of the inversions getting warmer. The inversions do still show a slight decrease in height as \textit{L}\textsubscript{\scriptsize{S}} progresses, and a reduction in the magnitude of the warming at the top of the inversion, but much less than in the in-track data. The inversions reappear a few degrees of \textit{L}\textsubscript{\scriptsize{S}} after the peak of the C storm. After this, inversions at this height do not occur again until the next large-scale dust storm in the following year.

The behavior of the ice clouds depends on the season. Prior to the A storm, the clouds increase markedly in opacity with local time. This is most notable at \textit{L}\textsubscript{\scriptsize{S}} $\approx$ 180\degree{}, where the ice opacity per kilometer increases by a factor of 10 between 2 am and 4 am (Figure \ref{fig:tdi_ls_alt}b,d,f). After the A storm, the clouds associated with the inversions show much less variability with local time, and overall tend to weaken between 3 am and 4 am. (An exception to this is at \textit{L}\textsubscript{\scriptsize{S}} $\approx$ 350\degree{}, where opacities again increase with local time). This seasonal and local time variability also occurs in the other Mars years with cross-track data available. This cloud behavior may explain why the inversions after the early MY29 Z storm are much stronger than those observed at any other time. The thick clouds are providing radiative heating during the day, and radiative cooling during the night. As the clouds are getting thicker as night progresses, they provide increasing radiative cooling, resulting in further cloud formation and a constructive feedback. This can strengthen the diurnal and semi-diurnal tides \citep{Hinson2004, Kleinbohl2013}, leading to increased dynamical warming at the tops of the inversions. Indeed, the lapse rates at the tops of the inversions follow the adiabatic lapse rate, suggesting warming via downwelling \citep{Steele2021}. Conversely, the lower opacity clouds which occur after the A dust storms tend to weaken as night progresses, and cannot cause as much strengthening of the diurnal and semidiurnal tides.

\begin{figure}[p]
  \includegraphics[width=1.0\linewidth]{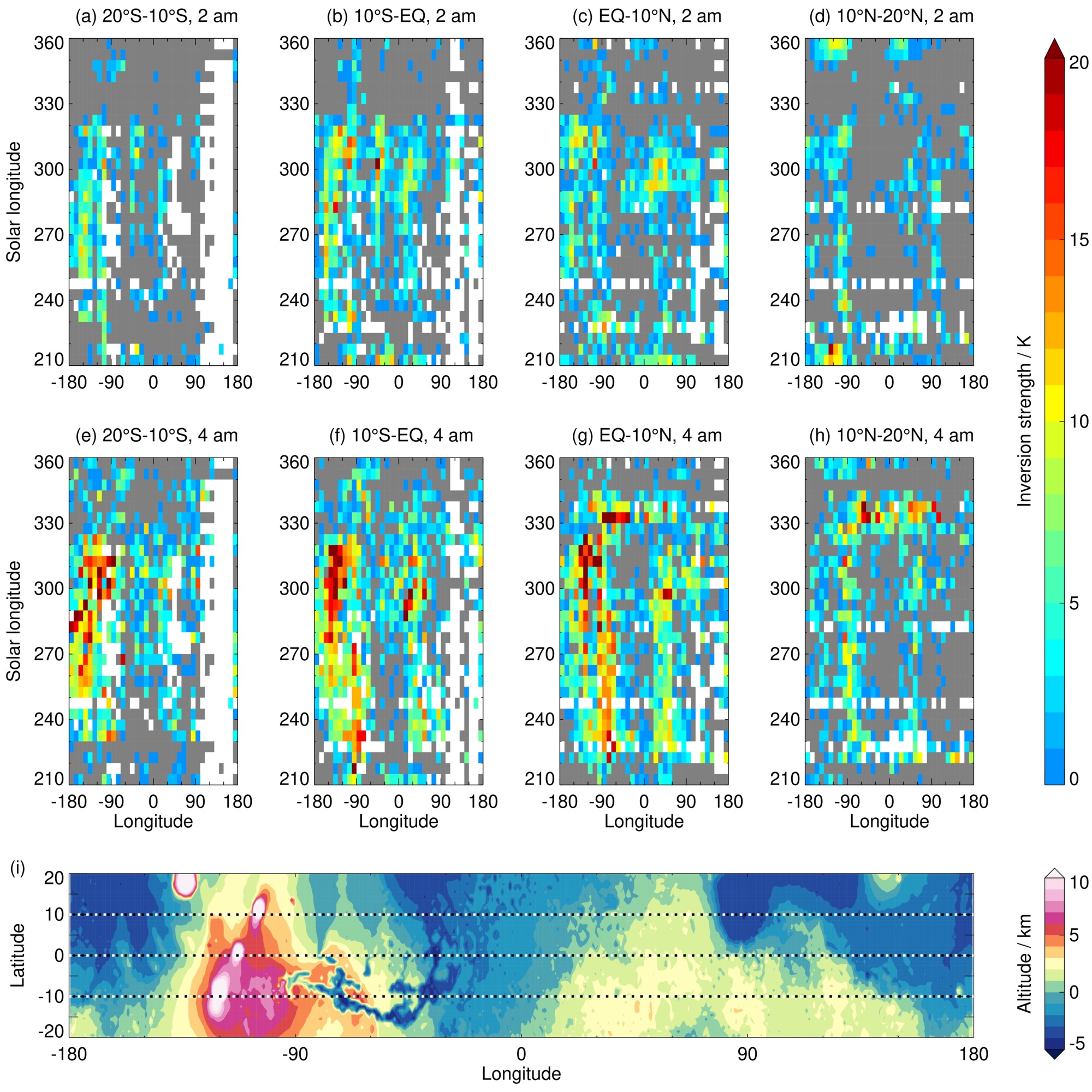}
  \caption{Location and strength of temperature inversions in MY33, as a function of longitude and \textit{L}\textsubscript{\scriptsize{S}}, at (a--d) 2 am and (e--h) 4 am. White regions show where no data are available, and gray regions show where data are present but no temperature inversions are observed. For each local time, four different latitude ranges in the tropics are shown. The topography of these latitude ranges can be seen in panel (i), which shows altitudes above the areoid from the Mars Orbiter Laser Altimeter \citep{Smith1999}.}
  \label{fig:invstrength}
\end{figure}

So far, our results have focused on temperature inversions located only in the 0--10\degree{}N and 60--90\degree{}W region, as these tend to be the longest lasting. However, inversions after dust storms occur at a variety of locations in the tropics. In order to show this geographic distribution, we calculate the inversion strength, which is the difference between the warmest temperature at the top of the inversion, and the coldest temperature at the base of the inversion. This was performed for all the temperature profiles at local times of 2 am and 4 am, and then the data were binned by 10\degree{} in latitude and longitude, and 5\degree{} in \textit{L}\textsubscript{\scriptsize{S}}. Figure \ref{fig:invstrength} shows how the inversion strengths at different latitudes in MY33 vary with local time, as a function of longitude and \textit{L}\textsubscript{\scriptsize{S}}. (Figures S4--S6 in the supporting information showing similar plots for MY30--32, while Figures S7--S10 show the 4 am inversion strengths for MY30--33 as a function of latitude and longitude.) The general increase in the strength of the inversions with local time is evident. The only region where the inversions weaken between 2 am and 4 am is at 10--20\degree{}N and \textit{L}\textsubscript{\scriptsize{S}} $\approx$ 350\degree--360{} (Figure \ref{fig:invstrength}d,h). As can be seen in Figure \ref{fig:invstrength}i, these inversions are located between Olympus Mons and Ascraeus Mons, and over Arabia Terra, with the same distribution also found at \textit{L}\textsubscript{\scriptsize{S}} $\approx$ 180\degree. These inversions occur lower in the atmosphere, with their bases at $\sim$100 Pa, as opposed to $\sim$20 Pa elsewhere.

When the inversions first form after the peak of the MY33 A storm, at \textit{L}\textsubscript{\scriptsize{S}} $\approx$ 220\degree{}, there are broad regions of inversions over Tharsis and Terra Sabaea, centered at around 90\degree{}W and 45\degree{}E (Figure \ref{fig:invstrength}f,g). These are regions of higher topography, and it is likely that the inversions are being caused by cloud radiative cooling coupled with the strengthened non-migrating thermal tides resulting from the increased dustiness \citep{Hinson2004}. However, there is also a region of inversions stretching across Chryse Planitia and Arabia Terra in the northern tropics (Figure \ref{fig:invstrength}g,h and S10). This region of inversions is even more extensive in MY32, which experienced a stronger A storm (Figures S6 and S9). At this time, this region does have a large abundance of dust associated with the dust storm, so the inversions are likely linked to tidal forcing caused directly by the dust, rather than topographically-excited tides. As the dust spreads across the globe and the dustiness decreases, the inversions over Chryse Planitia and Arabia Terra disappear, but remain over the Tharsis and Terra Sabaea regions. During this time, there is also a general trend for the inversions to extend to the west of Tharsis, into Amazonis Planitia. This is most notable in Figure \ref{fig:invstrength}f, where the inversions are strongest in this region between \textit{L}\textsubscript{\scriptsize{S}} $\approx$ 280--320\degree{}. Analysis of MCS dust profiles (not shown) show that dust in this region builds up between \textit{L}\textsubscript{\scriptsize{S}} $\approx$ 210--280\degree{}, and then remains constant until it decreases at \textit{L}\textsubscript{\scriptsize{S}} $\approx$ 320\degree{}. Thus, this westward shift of inversions is related to the build up of dust in Amazonis Planitia, which likely strengthens the diurnal and semidiurnal tides \citep{Hinson2004, Kleinbohl2013}. After the C dust storm at \textit{L}\textsubscript{\scriptsize{S}} $\approx$ 330\degree{}, inversions once again become prevalent in the northern tropics over Chryse Planitia and Arabia Terra (Figure \ref{fig:invstrength}g,h and S10). The same is true in MY32 (Figure S6), but in this year the C storm was weak, and the inversions are not as prominent. By \textit{L}\textsubscript{\scriptsize{S}} $\approx$ 345\degree{}, the inversions at these high altitudes across most regions have disappeared, and won't return again until the next large-scale dust storm.

\section{Conclusions}

Prominent nighttime temperature inversions have been observed in MCS data. They are ubiquitous in the dusty season, and are linked to the decay phases of dust storms. The bases of these temperature inversions typically occur at $\sim$20--30 Pa, with their tops at $\sim$5--7 Pa. These are higher in the atmosphere than those observed during the colder aphelion season \citep{Magalhaes1999, Hinson2004, Hinson2014, Wilson2014}. The strength of the inversions appears somewhat correlated with the strength of the dust storms, particularly for the A storms, with stronger inversions appearing after stronger storms. However, the inversions after the global dust storm in MY34 are no stronger than those after the A and C storms. When analyzing MCS data at different local times, the temperature inversions can be seen to strengthen between 2 am and 4 am, with the bases of the inversions getting cooler, and the tops of the inversions getting warmer.

Throughout most of the dusty season, the inversions are strongest in two main regions: Tharsis and Terra Sabaea. These are regions of higher elevation, and the MCS data show water ice clouds are present. Thus, it is likely that the inversions are forming due to the combination of stronger non-migrating thermal tides and cloud radiative cooling, as suggested for the inversions during aphelion season \citep{Hinson2004}. However, the locations of the inversions does change during the evolution of the dust storms. For example, during the decay phase of the A storms, strong inversions begin forming to the west of Tharsis, in Amazonis Planitia. MCS data suggest that these inversions are a response to the increased dust abundance in this area, as the dust from the A storm spreads across the globe. Thus the dust is likely causing a strengthening of the tides, and increased adiabatic warming of the descending air \citep{Hinson2004, Kleinbohl2013}. During the initial decay phases of dust storms, inversions are also located over the flat terrain of Chryse Planitia and Arabia Terra in the northern tropics. These inversions are likely due to the increased dust abundance strengthening the tides, as opposed to topographically-excited tides.

The results presented here show how varied the nighttime atmosphere in the dusty season can be, and how there is an important interplay between the dust distribution, water ice clouds and thermal tides. The diurnal and seasonal behavior revealed in these MCS observations is not well represented in assimilated datasets of the martian atmosphere \citep{Greybush2019, Holmes2020} and suggests further work is needed in order to understand the coupling between the dust and water cycles.

\section*{Acknowledgments}

The Mars Climate Sounder data used here are freely available from NASA's Planetary Data System at \url{https://atmos.nmsu.edu/data_and_services/atmospheres_data/MARS/mcs.html}. This work was performed at the Jet Propulsion Laboratory, California Institute of Technology, under a contract with NASA. Copyright 2021, California Institute of Technology. Government sponsorship acknowledged.

\bibliography{manuscript}

\end{document}